# A Provenance-Policy Based Access Control Model for Data Usage Validation in Cloud

Muralikrishnan Ramane[1], Balaji Vasudevan[2] and Sathappan Allaphan[3]

[123]Department of Information Technology, University College of Engineering
Villupuram, Tamilnadu, India
[1]murali.itpro@gmail.com, [2]jvbalajiit@gmail.com and [3]sathappan93@gmail.com

## ABSTRACT

*In an organization specifically as virtual as cloud there is need for access control systems to constrain users direct or backhanded action that could lead to breach of security. In cloud, apart from owner access to confidential data the third party auditing and accounting is done which could stir up further data leaks. To control such data leaks and integrity, in past several security policies based on role, identity and user attributes were proposed and found ineffective since they depend on static policies which do not monitor data access and its origin. Provenance on the other hand tracks data usage and its origin which proves the authenticity of data. To employ provenance in a real time system like cloud, the service provider needs to store metadata on the subject of data alteration which is universally called as the Provenance Information. This paper presents a provenance-policy based access control model which is designed and integrated with the system that not only makes data auditable but also incorporates accountability for data alteration events.*

## 1. INTRODUCTION

Cloud computing paradigm, [1] an internet technology that came into existence a decade ago is a boost to the world of resource sharing especially data resources. Cloud presents a demand-supply model for the services provided over the internet and some of the internet giants like Google, Amazon, Yahoo, .etc who have successfully implemented cloud computing models are still facing issues in effectively managing data residing in cloud. In a vastly distributed environment like cloud, putting data integrity into operation is intricate further to validate each and every event taking place is itself a huge workload for cloud service providers.

Cloud outsourcing [2] of data being processed is one the foremost reason for such uncertainty among users who utterly trust the cloud for storing their private data. Still the major competitors in the market are providing various conventional security and access control schemes like privacy preserving and protection, public verifiability, auditability and accountability do not allay user's concerns on handling their personally identifiable information. To deal with such issues, cloud needs some access control schemes that not only allow users to audit their private data but also track its usage.

In cloud, data auditing and accounting [3] are tied together since auditing involves tracking and logging data usage events and consequently auditing the data usage logs helps to find the user accountable for the violation. In this milieu the provenance information (PI) plays a major role towards data integrity management which is in the form of metadata, the history detailing the derivation of the objects and contains information that allows policy-independent access control decisions.

An important requirement of any information management system is to protect data and resources against leak or improper modifications, while at the same time ensuring their availability to legitimate users. Enforcing protection therefore requires that every access to a system and its resources be controlled and this goes under the name of access control. Access Control is considered as one of critical security mechanisms for data protection in cloud applications. Unfortunately, traditional data access control schemes usually assume that data is stored on trusted data servers for all users and this assumption no longer holds in cloud computing since the data owner and cloud servers are very likely to be in two different domains.

The access control mechanism [4] as shown in Figure 1; must work as a reference monitor, a trusted component intercepting each and every request to the system. It must also enjoy the following properties such as tamper-proof, non-bypassable and secured kernel. Access control policies [5] can be grouped into three main classes: Discretionary Access Control (DAC) policies control access based on the identity of the requestor and on access rules stating what requestors are allowed to do. Mandatory (MAC) policies control access based on mandated regulations determined by a central authority. Role-based access control (RBAC) policies control access depending on the roles that users have within the system and on rules stating what accesses are allowed to users in given roles.

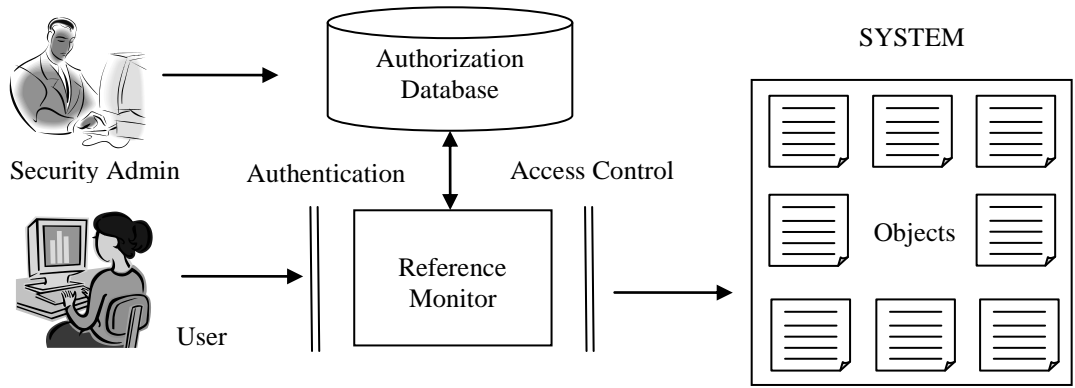

Figure 1. Fundamental Access Control Mechanism

Provenance based access control [6] is a recent trend in organizations where user's private data must be protected. It provides additional capabilities to service providers unlike conventional access control models like DAC, MAC, RBAC and well suited than ABAC and HBAC models. The fundamental focus of this work is on how to enhance the security in the system using provenance information which entirely relies on security policies and roles enforced on the system and identify whoever violates the protocols for data usage. This paper contributes to access control schemes and enforces a novel security technique on data storage in cloud. This paper proposes a provenance based access control scheme which is fundamentally a PBAC model. The rest of this paper is structured as follows; Section 2 summarizes the related work on previous access control techniques. Section 3 proposes the basic design for the model. Section 4 concludes the paper and suggests future work.

## 2. RELATED WORK

Cloud computing enables highly scalable services consumed over the internet on an as-needed basis. A major feature of the cloud services is that user's data are usually processed remotely in unknown machines that users do not own or operate. In recent years fear of losing control over their own data has become a significant barrier to the wide adoption of cloud services. To address this problem, several authors have provided solutions in their own style. There are several systems designed to secure data using provenance and similarly to secure provenance information. The work on accountability for data sharing in cloud is the base of the proposed model. The main contribution of our work is developing an access control policy for data based on provenance records collected from provenance store in cloud thereby enhancing security. The following are few of the solutions for the above said problem.

## 2.1. Provenance Based Access Control Model

In this paper [7] the author proposes a novel provenance-based access control model that addresses the objective of how provenance data can be used to enhance security in the system. Using provenance data for access control to the underlying data facilitates additional capabilities beyond those available in traditional access control models. This paper utilizes the notion of dependency as the key foundation for access control policy specifications. This model can support dynamic separation of duty, workflow control, origin-based control, and object versioning. The proposed model identifies essential components and concepts and provides a foundational base model for provenance-based access control.

## 2.2. CloudPolice - Taking Access Control out of the Network

This paper [8] argues that it is both sufficient and advantageous to implement access control only within the hypervisors at the end-hosts because Access control techniques are intricate due to multi-tenancy, the growing scale and dynamicity of hosts within the cloud infrastructure, and the increasing diversity of cloud network architectures. The author proposes Cloud-Police, a system that implements a hypervisor-based access control mechanism which is simpler, more scalable and more robust than existing network-based techniques.

## 2.3. Secure and Efficient System for End-to-End Provenance

EEPS [9] collects provenance evidence at the host level by trusted monitors. Provenance authorities accept host-level provenance data from validated monitors to assemble a trustworthy provenance record. EEPS addresses the critical open problem of showing that provenance information was recorded accurately within and across systems EEPS introduces the notion of a host-level provenance monitor. The author proposes the host level provenance monitor as a method for achieving these guarantees. The system uses the notion of a plausible history as a method for tracking a data item's history of domain traversals. Finally the provenance information are collected, stored, and audited with optimized cryptographic constructions for provenance data.

## 2.4. Accountability for Data Sharing in Cloud

In cloud user's data are usually processed remotely in unknown machines that users do not own or operate which can become a significant barrier to the wide adoption of cloud services. To address this problem [10] a novel highly decentralized information accountability framework is designed to keep track of the actual usage of the users' data in the cloud. The author proposes an object-centered approach that enables enclosing our logging mechanism together with users' data and policies and ensures that any access to users' data will trigger authentication and automated logging local to the JAR.

## 3. PROPOSED WORK

This work attempts to develop an access control policy based on provenance records collected from provenance store in cloud. Every access control system [11, 12] has three levels of abstraction such as Policy, Mechanism, and Model. An access control policy authorizes a set of users to perform a set of actions on a set of resources within an environment. Unless authorized through one or more access control policies, users have no access to any resource of the system. Access control policies can be grouped into three main classes such as RBAC, DAC and MAC which restricts access based on roles, identity of the user, and mandated regulations determined by a central authority respectively.

In RBAC each time a user does not have access to an object through an existing role, a new role is needed. As the policies become more fine-grained, a role is needed for each combination of the different resources in the provenance graph. Similar drawbacks apply to the DAC and MAC access control models since they both use mapping functions to associate users with objects. Clearly, applying these traditional access control policies for fine-grained access control in provenance would result in prohibitive management costs. Moreover, their usage in provenance would be an arduous task for the administrator.

Provenance is represented by a set of records stored in a provenance store. Basically there are five kinds of provenance records, such as Operation records, Message records, Actor records, Preference records and Context records. Each provenance record is identified uniquely by ID attribute and a Timestamp attribute to identify version of provenance. The schema of a provenance record is shown in Figure 2; Operation record includes ID, actor ID, context ID, description, output, and timestamp. Message record includes ID, actor ID, source ID, destination ID, Description, Content carrier an Timestamp. Actor records include ID, Name and Role. Context record includes ID, State and parameter. Preference record includes ID, Target, Condition, Effect, Obligations and Timestamps.

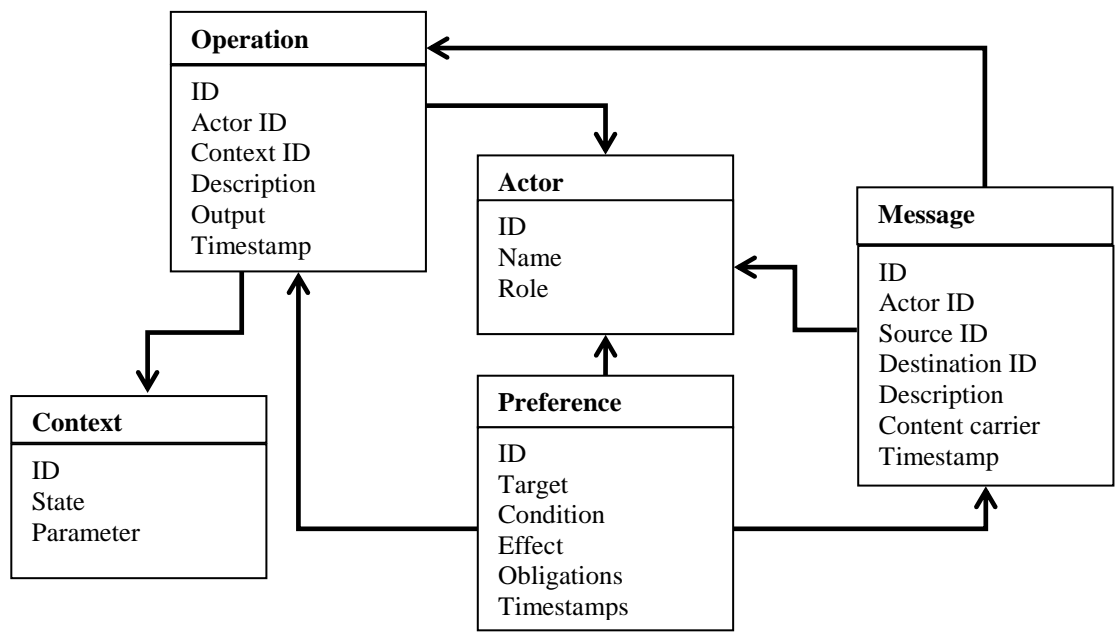

Figure 2. Schema of a Provenance Record

## 3.1. Provenance Based Access Control

The model is designed with a notion to enhance security in a cloud organization which adopts Cloud Information accountability framework for providing data integrity of user data. Users in cloud come in two types, Authenticated and Unauthenticated. This model doesn't deal with the problem of unauthenticated users having access into the system, instead it handles data corruption problem by an authorized user who in the process of accessing the information may corrupt it. This model cures the system from data corruption rather preventing it using a provenance based access control model. The model analyses user behavior using provenance records and generates policies that are attached with a data resource before uploading it to the cloud.

### 3.1.1. Introduction

The model as shown in Figure 3; is designed using four components and two fundamental operations: Data owner, Cloud, CIA Framework & CSP and Policy generation and Policy evaluation respectively.

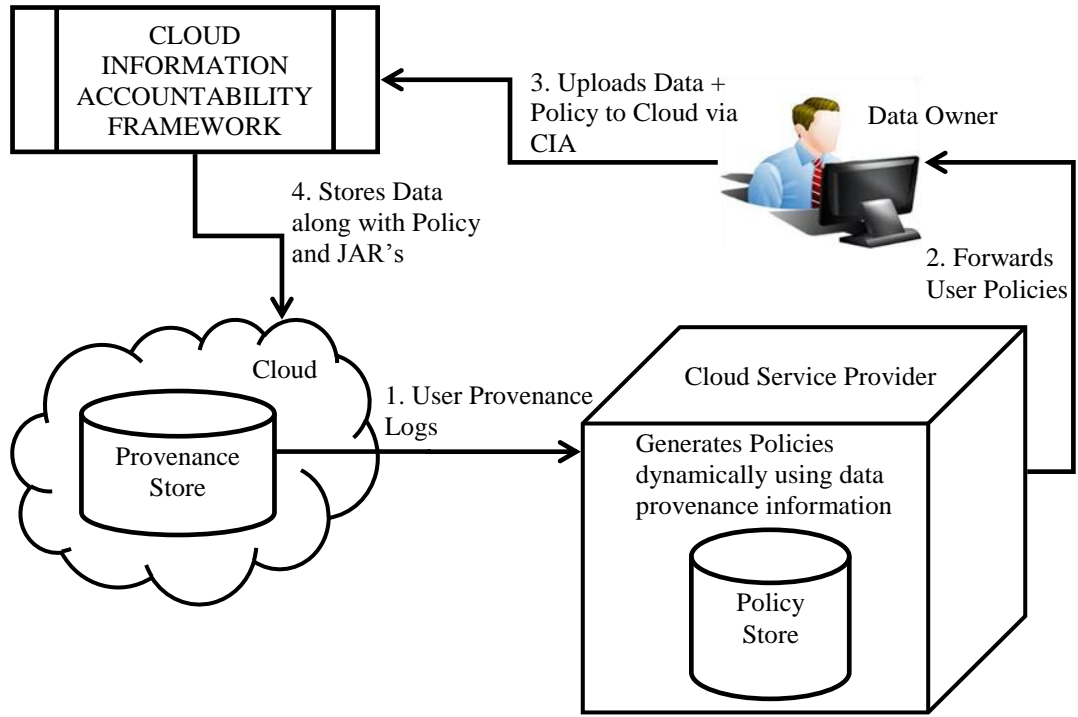

Figure 3. Provenance-Policy Based Access Control Model

- Data Owner

Data owner is the one uploads the data resource to the cloud and the one who has made an agreement with cloud service provider in terms of privacy, integrity and data loss. The owner accepts generated policies from the CSP and attaches it with every data resource that uploaded into the cloud. The owner trusts the CSP upon the policies generated so that the data couldn't be altered or corrupted by other users the cloud.

- Cloud service Provider

The CSP is the major component in the model that generates policies based on provenance information collected from provenance store. The policies are generated based on provenance records which provide details on User ID, Name, Role, Timestamp, context, etc. on users accessing the data resource.

- CIA Framework

This framework is adopted from previous work on data accountability in cloud where it generates jar's which are similar to digital signatures and attaches them with every data being uploaded to the cloud. The jar's provide a form of security which gets corrupted or invalid if there is any illegal or incorrect access to the data resource. Additionally in this paper it is redesigned to attach provenance policies with every data resource being uploaded to cloud.

- Cloud

Cloud basically provides the storage area for huge amount of data resources that are shared by users and organizations. In this model it additionally hosts a provenance store that collects provenance information from every data access request made to the cloud. The provenance information is stored in the form of provenance records which is input to the policy generation functionality.

### 3.1.2. Policy Generation

In this phase policies are generated using provenance records and its attributes are analyzed to gather details on following:
- Which user is accountable for the data alteration?
- In which role the corresponding user was access the resource?
- At what time instant the resource was accessed?
- Under what context the resource was accessed?

To answer the above questions a separate policy record is generated by aggregating operation, message, preference, actor and context records as shown in figure 4.

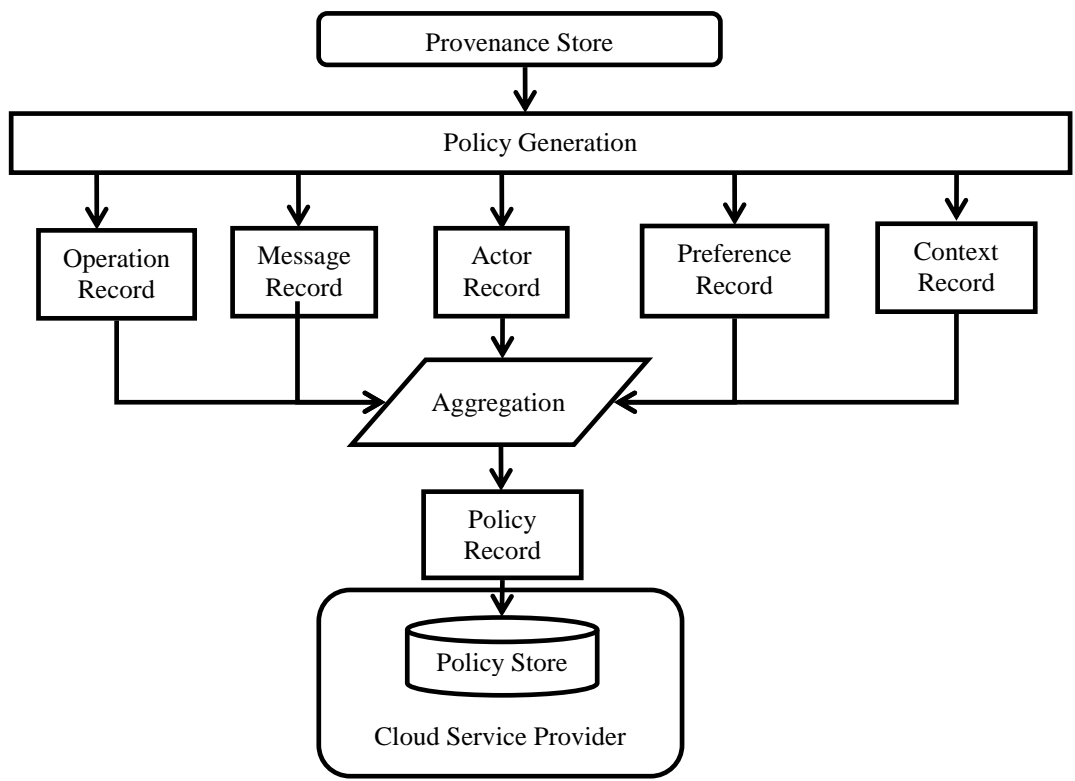

Figure 4. Flow Diagram for Policy Generation

A sample policy attached with a data resource:

```
<policy ID="1" >
<target>
<subject> Actor.ID </subject>
<record>Operation.description</record>
<restriction>Actor.role=="AuthorizedUser"</restriction>
</target>
<condition> system.machineid == "192.168.2.35" </condition>
```

```
<effect> Permit </effect>
<obligation>
<temporal constraint> 10 days </temporal constraint>
</obligation>
</policy>
```

The policy record is generated by selecting particular attributes from other provenance records which is necessary to identify a user; in which role; under what context the resource was accessed. The major contribution is given by operation record which provides details on all the data access transactions taken place in cloud.

| Policy Record |
| --- |
| Operation.Actor ID<br>Operation.Context ID<br>Operation.Timestamp<br>Operation.Actor ID $\cap$ (Actor.ID $\times$ Actor.Role) |

### 3.1.2. Policy Evaluation

In this phase every data access request is analyzed by the evaluator for the requestor behavior. The generated policy record is queried to find that the corresponding user, using present role, under same context have never misbehaved. The requests that are evaluated to success leads to partial or full access of the data resource requested.

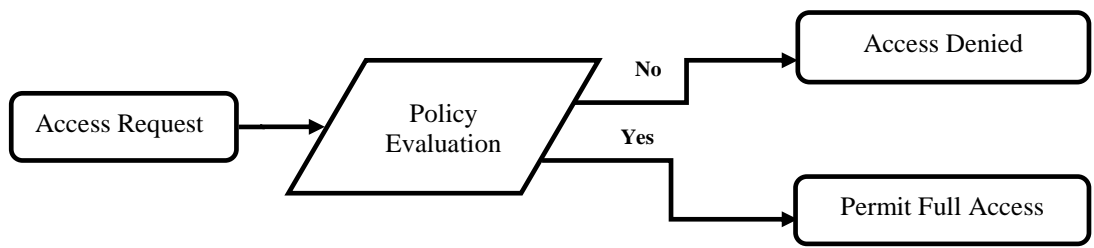

Figure 5. Flow Diagram for Policy Evaluation

## 4. CONCLUSION AND FUTURE WORK

To conclude, the proposed system generates a security policy that significantly depends on the provenance information of the particular data object which is subsequently attached to the data before storing it in a public cloud. This work relies on the dynamic policies generated by the service provider which incorporates access control into the system. Further the system is in the process of testing within a small private cloud which contains few hundreds of data. The policies generated currently provides limited access control on users data and could be extended in terms of dynamic and swift policy generation techniques that improve the performance of the system. The cloud is benefited the most as the proposed access control model not only makes data auditable but also incorporates accountability for data alteration events. This model could be incorporated in any auditing scheme as an additional support that cures the cloud system from misbehaving and mishandling users.

## ACKNOWLEDGEMENTS

The authors would like to thank all those who have shared their valuable inputs, insights, suggestions and for their time throughout the course of this work.

**Authors**

Muralikrishnan Ramane has completed post graduate degree in the field of Computer Science Engineering, specialized in Distributed Computing Systems. He is currently working as a Lecturer at University College of Engineering Villupuram, Tamilnadu, India. To his credit he has three International Journal and one International Conference publications. His research interests include Data Management in Distributed Environments.

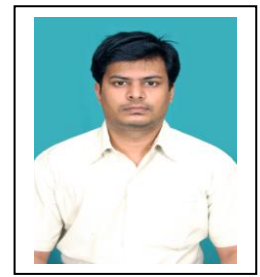